\documentclass[12pt]{article} 
  
\usepackage{epsfig}
\usepackage{calrsfs} 
\usepackage{bbm,bm}
\usepackage{cite}
\usepackage{color,cancel,graphicx,mathrsfs}
\usepackage{amsmath}
\usepackage{mathtools}
\usepackage{nicefrac}
\usepackage{slashed}
\usepackage{unitsdef}
\numberwithin{equation}{section} 

\usepackage{exscale}

\usepackage{subcaption}

\usepackage{hyperref}
\usepackage{xcolor}
\hypersetup{colorlinks, linkcolor={red!50!black}, citecolor={blue!50!black}, urlcolor={blue!80!black}}

\def\a0size{6}

\newcommand{\cw}{c _ {  W }} 
\newcommand{\Gsph}{\Gamma _ {\rm sph} }

\newcommand{\fB}{f _ { \rm B } } 
\newcommand{\fF}{f _ { \rm F } } 
\newcommand{\gs}{g_S} 
\newcommand{\gw}{g_W} 
\newcommand{\aw}{\alpha _ W} 
\newcommand{\gvec}[1]{\mbox{\boldmath $#1$}}
\newcommand{\mmq}{m ^ 2  _ { { \infty } }}

\newcommand{\mmD}{m ^ 2  _ { G }}

\newcommand{\mD}{m  _ { G }}
\newcommand{\mmW}{m^2  _ { {  W} }}
\newcommand{\mW}{m  _ { W }  }
\newcommand{\mmWq}{m^2  _ {   W  q }}
\newcommand{\mWq}{m  _ { { \rm W } q }}
\newcommand{\nc}{N _ { \rm c}}
\newcommand{\nfam}{N _ { \rm fam }}

\newcommand{\rmi}[1]{{\mbox{\scriptsize #1}}}
\newcommand{\rscale}{\mu}
\newcommand{\betas}{\beta_S}
\newcommand{\betaw}{\beta_W}
\newcommand{\mz}{m_Z}
\newcommand{\W}{A}

\newcommand{\lsi}{\raise0.3ex\hbox{$<$\kern-0.75em\raise-1.1ex\hbox{$\sim$}}}
\newcommand{\gsi}{\raise0.3ex\hbox{$>$\kern-0.75em\raise-1.1ex\hbox{$\sim$}}}
\newcommand{\lsim}{\mathop{\lsi}}

\renewcommand{\bm}[1]{{\bf #1}}
\renewcommand{\vec}[1]{{\bf #1}}

\begin{document} 

\setlength{\baselineskip}{0.6cm}
\newcommand{\figysize}{16.0cm}
\newcommand{\figtopspace}{\vspace*{-1.5cm}}
\newcommand{\figbottomspace}{\vspace*{-5.0cm}}
  
\newcommand{\db}[1]{\textcolor{blue}{#1}}
\newcommand{\pk}[1]{\textcolor{orange}{#1}}
\newcommand{\pkcom}[1]{\textcolor{orange}{[PK: #1]}}
\newcommand{\dbcom}[1]{\textcolor{blue}{[DB: #1]}}

\begin{centering}
\vfill

{ 
\centerline{ \Large \bfseries
QCD corrections to the electroweak sphaleron rate
    }
}

\vspace*{1cm}

Dietrich B\"odeker$^\rmi{a,}$\footnote{\texttt{bodeker@physik.uni-bielefeld.de} },
Philipp Klose$^\rmi{b,}$\footnote{\texttt{pklose@nikhef.nl}}

\vspace*{.6cm}

$^\rmi{a}$%
{\em 
Fakult\"at f\"ur Physik, Universit\"at Bielefeld, 33501 Bielefeld, Germany
}

\vspace*{0.3cm}

$^\rmi{b}$%
{\em
Theory Group, Nikhef, Science Park 105, 1098 XG Amsterdam, The Netherlands
}

 
\end{centering}

\begin{abstract}
\noindent
The electroweak sphaleron rate in the high temperature phase of the 
Standard Model
is inversely proportional to the weak-isospin conductivity.
So far, only electroweak interactions were included in its computation.
Here we take into account quark scattering through strong interactions 
at leading-log order.
These reduce the quark contribution to the conductivity
by up to 15\%, and the total conductivity by up to 6\%.
\end{abstract}

\newpage
\tableofcontents
 
\section{Introduction}  
\label{intro} 

The rate for electroweak baryon 
number violation at finite temperature $ T $,
also called the sphaleron rate, is a key ingredient in scenarios
that could explain the baryon asymmetry of the Universe.%
\footnote{ For a recent review, see e.g. \cite{Bodeker:2020ghk}.} 
At temperatures exceeding 
the electroweak phase transition (or crossover) temperature
$ T _ c $,
the rate is larger than the Hubble rate,
leading to rapid baryon number violation.
Around $ T _ c $, the rate freezes out, so that
baryon number is conserved from then on. 
At higher temperatures, the rate
is determined by the non-perturbative 
dynamics of electroweak gauge bosons
with wavelengths  of order of the electroweak magnetic screening length 
$\ell_\text{mag} \sim ( \gw ^ 2 T ) ^ { -1 }  $,
where $ \gw $ is the weak-SU(2) gauge coupling.
Due to their large occupation number $ \sim T/| \vec k | \sim \gw ^{ -2 } $,
they can be treated as classical fields.
Their dynamics is 
determined by the non-Abelian 
counterpart of Ampere's law
\begin{align}
   \vec D \times \vec B = \sigma  \vec E + \gvec { \zeta }
   \, .
   \label{langevin} 
\end{align} 
The right-hand side of eq.~\eqref{langevin} is 
the weak-isospin current, which 
captures the medium response due to particles
with hard momenta
$ | \vec p | \sim T $.
The ohmic part of the current is local in the weak-electric field 
$ \vec E $ because the 
mean free path of isospin-changing small-angle scattering 
is parametrically of order
$ [  \gw ^ 2 \log ( \gw ^{ -1 } )  T  ] ^ { -1 } $.
In a leading log approximation, this length scale
is much smaller than the 
magnetic screening length.
The coefficient
$ \sigma  $ is the weak-isospin conductivity,%
\footnote{In the context of QCD it is referred to as  color conductivity.}
and $ \gvec \zeta  $ is a Gaussian white noise term generated by thermal 
fluctuations of the high-momentum modes.

The only length scale in this equation is $ \ell_\text{mag} \sim ( \gw ^ 2 T ) ^{ -1 } $.
We can therefore estimate $ \vec D \times \vec B \sim \ell_\text{mag}^{ -2 } \vec \W $,
where $ \vec \W $ is the weak SU(2) vector potential.
The first term on the right-hand side of \eqref{langevin} acts like a 
damping term for the gauge fields,
and determines
the characteristic time scale $ \tau  $
of the isoelectric field $ \vec E \sim \tau  ^{ -1 } 
\vec \W $ as $ \tau \sim \sigma \, \ell_\text{mag}^2 $.
The value of the conductivity is large, 
$ \sigma  \sim T /\log (\gw ^{ -1 } ) $,
which leads to over-damping of the gauge fields.
Therefore, eq.~\eqref{langevin} does not contain a 
``displacement current'' $D_t \vec E$,
which would be of order $ \tau  ^{ -2 } \vec \W $.

The hot electroweak sphaleron rate $ \Gsph $, 
i.e. the rate of 
sphaleron transitions per unit time and unit volume, 
is parametrically of order 
$\Gsph \sim ( \tau \ell _ \text {mag}  ^ 3 ) ^{ -1 } $
and can be written as \cite{Bodeker:2020ghk}  
\begin{align}
    \Gsph
     = k  \frac { 2 \pi T } { 3 \sigma  } \aw  ^ 5 T ^ 4 
   \label{Gsph} 
\end{align}
with $ \aw \equiv \gw ^ 2/(4 \pi  ) $. 
The dimensionless number 
$ k $  
has been determined through  lattice simulations of eq.~(\ref{langevin})
\cite{Moore:1998zk,Moore:2010jd},%
\footnote{Close to the electroweak crossover one needs to
include the Higgs field in the effective infrared theory \cite{Moore:2000mx},
for recent results, see \cite{ Annala:2023jvr}.}
  yielding
\begin{align}
    k  =  10.0 \pm 0.3 
   \label{kappa} 
   \, .
\end{align}

So far, only electroweak interactions were included in the computation
of $ \sigma  $.
However, almost half of the weak-isospin current is carried by quarks,
which also undergo scattering through strong interactions.
For these, the mean free path is 
of order $ [ \gs ^ 4 \log (\gs ^{ -1 } ) T ] ^{ -1 } $.
Counting $ \gs \sim \gw $ this would be much larger than the mean
free path $[ \gw^2 \log(\gw^{-1}) T ]^{-1}$ for electroweak scattering
and thus negligible.
The reason for this difference is that 
scatterings of hard particles that are mediated by $ W $-boson exchange 
are enhanced for small scattering angles $ \gw ^ 2 \lsim  \theta  \lsim  \gw $.
This process hardly changes the momentum of the scattered particles,
but nevertheless does impact the weak-isospin current because it \emph{does} 
change their weak isospin.
In contrast, quark-quark scattering 
via gluon exchange does 
not change the weak isospin of the quarks,
so that the weak-isospin current is only affected
by changing the momentum of the particles.
Correspondingly, the scattering is dominated by angles
$ \gs  \lsim  \theta  \lsim  1 $.
The same reasoning applies to QCD Compton scattering
and quark-antiquark annihilation with small momentum exchange.
While these processes do change weak isospin,
their matrix elements are less singular, so that scattering with angles less 
than $ \gs $ are suppressed by QCD Debye screening \cite{Arnold:1999uy}.

However, the strong coupling constant $\gs$ is numerically larger than $\gw$,
which may partially compensate the small-angle enhancement of the electroweak scattering.
The main aim of this work is therefore to determine the quantitative 
impact of QCD corrections
of the quark contribution to the 
weak-isospin conductivity 
$ \sigma  $.
To incorporate the effect of QCD scattering on the electroweak sphaleron rate,
it suffices to insert the new value of $ \sigma  $ in eq.~\eqref{Gsph}, 
without altering the value of $ k $.

The remainder of this work is organized as follows:
In sec.~\ref{sec:vlasov} we review the Vlasov equations of electrodynamics
describing the hard thermal loop effective theory,
their non-Abelian generalization
and the resulting Boltzmann equation containing the
electroweak collision term.
In sec.~\ref{sec:qcd coll}, we compute the QCD collision term.
Sec.~\ref{sec:numerics} presents the numerical inversion of
the complete quark collision 
term to obtain the corrected isospin conductivity,
and sec.~\ref{s:concl} concludes the paper.
This paper includes three appendices. 
Appendix \ref{sec:couplings} collects
the coupling constants used to obtain
our numerical results. 
A derivation of the Vlasov-Boltzmann equations from Schwinger-Dyson equations is provided
by appendix.~\ref{sec:vlasov-boltzman derivation}.
Appendix  \ref{eq:SD collision term} details a derivation
of the QCD collision term from a
2-particle-irreducible (2PI) effective action.

\section{Vlasov-Boltzmann equation}  
\label{sec:vlasov} 

To begin, we recall the 
Vlasov equations that govern the dynamics of an electrodynamic plasma
and their relation to the physics of 
hard thermal loops (HTLs) in QED 
(cf.\ \cite{Silin:1960pya,Blaizot:1992gn}).
These equations describe the coupled evolution of
long wavelength ($ | \vec k | \ll T $)
thermal fluctuations of the 
electromagnetic field
and the phase space densities of charged particles
with hard ($ | \vec p | \sim T $) momenta
in a hot plasma. 
Such fluctuations are present even in thermal equilibrium.
Due to their large occupation number, the long-wavelength 
fluctuations of the electromagnetic field can be treated
as classical fields, while the charged particle are 
treated as classical particles.

The phase space densities $ f _ { e ^ \pm } ( t, \vec x , \vec p ) $
of positrons and electrons then
evolve according the Vlasov equations \cite{Lifshitz1995Physical}
\begin{align}
   (\partial_t + \vec v \cdot \nabla) f _ { e ^{ \pm } } 
   \pm e ( \vec E + \vec v \times \vec B ) 
   \cdot \frac { \partial f _ { e ^ \pm } } { \partial \vec p } 
   = 0 \, ,
   \label{vlasovqed} 
\end{align} 
where $ \vec v $ is  the particle velocitity
and $ e $ is the positron charge.
This equation neglects the effect of hard  
gauge fields, which would lead to an additional collision term.

We consider high temperatures, such that we can neglect the electron mass and
the velocity is given by $ \vec v = \vec p /| \vec p | $.
The electric current is
\begin{align}
   J_\text{em}^ \mu  = 2 e \int \frac  { d ^ 3 p } { ( 2 \pi  ) ^ 3 } 
   ( f _ { e ^ + } - f _ { e ^ - } ) v ^ \mu  \, ,
   \label{ecur} 
\end{align} 
where we have assumed that both spin polarizations give the same contribution,
resulting in the factor 2.
For small deviations from equilibrium 
$ f _ { e ^ \pm } = f _ { \rm F } + \delta  f_ { e ^ \pm } $,
where $ \fF (  \vec p ) \equiv  ( e ^{ | \vec p | /T } + 1 ) ^{ -1 } $ 
is the Fermi-Dirac distribution,
one may linearize 
eq.~(\ref{vlasovqed}).
This gives
\begin{align}
   (\partial_t + \vec v \cdot \nabla) \delta  f _ { e ^ \pm } 
   \pm e  \,\vec E \cdot  \vec v \,
   f' _ { \rm F }  
   &= 0 
   \label{vlasovlin} 
\end{align} 
with 
$    f' _ { \rm F } \equiv  \partial f _ {\rm F  }  / \partial | \vec p | $.  
By solving eq.~\eqref{vlasovlin} for $ \delta  f $, inserting the solution into the 
current~\eqref{ecur} and Fourier transforming, one obtains 
\begin{align}
   J_\text{em}^ \mu  (k ) = 
   \Pi  ^{ \mu  \nu  } _ { \rm HTL } ( k ) A _ \nu  ( k ) \, ,
\end{align}
where $ \Pi  ^{ \mu  \nu  } _ { \rm HTL } $ is the hard thermal loop
polarization tensor. This tensor is derived from thermal field theory
by integrating out hard modes to obtain an effective
theory for low-momentum $ k ^ \mu  \ll T $ modes of the 
electromagnetic field.
In essence, the physics of classical charged particles interacting
with a classical gauge field is same as the physics of hard thermal loops.
The HTL effective theory is then described by 
eqs.~\eqref{ecur} and \eqref{vlasovlin} together with 
Maxwell's equation \cite{Blaizot:1992gn} 
\begin{align}
   \partial _ \mu  F ^{ \mu  \nu  } = J_\text{em}^ \nu  \ .
   \label{emmaxwell} 
\end{align}

In the non-Abelian weak SU(2) theory,
there is not only
a HTL polarization tensor, but also HTL $ n $-point 
functions for all $ n > 2 $. 
Their generating functional is determined by
non-Abelian counterparts of eqs.~\eqref{ecur} and \eqref{vlasovlin} 
\cite{Blaizot:1994am},
with one distribution function for each isospin-charged particle in the 
Standard Model.
In the following, we focus on the quark contribution to 
the weak-isospin current
\begin{align}
   J ^ a _ { q \, \mu   } = \frac {\gw}2 \nfam \nc
   \int \frac { d ^ 3 p } { ( 2 \pi  ) ^ 3 }
   \left [ \delta  f ^ a _ + - \delta  f ^ a _ - \right ]   
   v _ \mu \ ,
   \label{qcur} 
\end{align}
where we have summed over colors and 
quark isospin doublets, resulting in the factors $ \nc = 3 $ 
and $ \nfam = 3 $, respectively.
Note that there is no spin sum, as only left-handed quarks and
right-handed antiquarks couple to the SU(2) gauge bosons.
Since the doublets transform under the fundamental representation of SU(2),
their distribution functions $ \delta  f ^ a _ \pm $,
defined as moments of quark two-point correlation functions,
transform under the adjoint representation.%
\footnote{See also the discussion in 
app.~\ref{sec:vlasov-boltzman derivation} for details.}
This behavior is similar 
to  lepton flavor oscillations,
which can be described by promoting the neutrino distribution function to a 
3$ \times $3 matrix in flavor space.
The non-Abelian version of eq.~\eqref{vlasovlin} for quarks 
is given by~\cite{Blaizot:1994am}
\begin{align}
  (D_t + \vec v \cdot \vec D) \delta  f _ \pm  
   \pm \gw \vec v \cdot \vec E \, \fF '= 0 
   \, ,
  \label{vlna} 
\end{align}
where $ D _ \mu  $ is the SU(2) covariant derivative in
the adjoint represenation.
The complete HTL effective theory
for gauge fields with small momenta $| \vec k |  \ll T$
is then fully described \cite{Blaizot:1993be} 
by eqs.~\eqref{qcur} and \eqref{vlna}
along with the Yang-Mills equation
\begin{align}
   D _ \mu  F ^{ \mu  \nu  } = J ^ \nu \ ,
   \label{ym} 
\end{align}
where the current $ J ^ \mu  $ is defined by adding the quark contribution \eqref{qcur}
to the analogous contributions from the lepton, $ W $ boson, and Higgs fields.
Ultimately, we want to describe
magnetic-scale gauge fields.
For these, the HTL polarization tensor
is of order $ \gw ^ 2 T ^ 2 k _ 0/| \vec k | $, which is much larger
than the tree-level terms in the propagator, unless $ k _ 0 \ll | \vec k | $.
In other words, they are strongly overdamped \cite{Arnold:1996dy}, 
rendering time derivatives in eqs.~\eqref{vlna} and \eqref{ym}
negligible. 

By further intergrating out fields with momenta 
$ k ^ \mu  \sim \gw T $,
one obtains an effective theory for the fields with momenta
on the order of the magnetic screening scale $ \gw ^ 2 T $. 
Their dynamics
is governed by equations of motion that take the same form as 
eqs.~\eqref{qcur}, \eqref{vlna}, and \eqref{ym},
except that there is now a gaussian white noise and a linear 
collision term  on the right-hand side of eq.~\eqref{vlna},
which we denote as $\left [ \dot f _ { \pm } \right ] _ { \rm coll }$.
The collision term is of order $ \gw ^ 2 \log(\gw ^{ -1 } ) T \delta  f  $, 
while the term $ \vec v \cdot \vec D \, \delta  f $  
on the left-hand side of eq.~\eqref{vlna} 
is of order  $ \sim \gw ^ 2 T \delta  f $.
In a leading-log approximation, which turns out to be valid also at
next-to-leading log order, one can 
therefore neglect the latter term, which yields 
\begin{align}
\label{eq:lw boltzmann}
\pm \gw \vec v \cdot \vec E \, \fF '= 
\left [ \dot f _ { \pm } \right ] _ { \rm coll } \ .
\end{align}
Here we have not displayed the white noise term that also
results from integrating out the modes with $ | \vec k | \gg \gw ^ 2 T $.
This term gives rise to the noise term in eq.~\eqref{langevin}, but it 
is not needed for our computation of the isospin conductivity.

The linearized collision term
can be viewed as a  linear
operator acting on $\delta f$.
This operator commutes with rotations acting on
the velocity variable $ \vec v $.
Thus its eigenfunctions are the spherical harmonics 
$ Y_ { lm} ( \vec v ) $, 
and its eigenvalues only depend on~$ l$~\cite{Bodeker:1999ey}.
Since the left-hand side of eq. \eqref{eq:lw boltzmann}
is a linear combination of spherical harmonics with $l = 1$, 
only the $ l=1 $ sector of $ \delta  f _ \pm $ is nonzero,
so that
\begin{align}
\label{eq:weak collision term}
\left [ \dot f _ \pm \right ] _ { \rm coll } &=
-
 \cw \, \delta f_\pm \ ,
\end{align}
where $ \cw$ is minus the $ l =1 $ eigenvalue
of the collision term.
At next-to-leading log order, it is given by \cite{Arnold:1999uy}
\begin{align}
   \label{c1w} 
\cw &=  \gamma 
+ 
 \frac{\gw^2 T}{2\pi} K \ , &
K &= 3.0410 \ ,
\end{align}
where $\gamma$ is defined implicitly as a solution of the equation
\begin{align}
   \gamma = \frac{\gw^2 T}{2\pi} \log\left( \frac{m_W}{\gamma} \right) 
\ .
\end{align}
Here  
\begin{align} 
\mmW = \left( \frac23 
   + \frac{N_s}6 + \frac{(1+\nc)\nfam}{12} \right) \gw^2 T^2 
   \label{mw} 
\end{align}
is the $ W $-boson Debye mass with $N_s = 1$ the number of Higgs doublets in the SM.
Table \ref{t:gamma} presents a selection of values for $ \gamma  $.
Using
eq.~\eqref{eq:weak collision term} in eq.~\eqref{eq:lw boltzmann},
one finds 
\begin{align}
\delta f  _ \pm &= \frac{
\mp
 \gw \vec v \cdot \vec E}{\cw} \fF' 
   \ .
\end{align} 
Inserting this solution into eq.~\eqref{qcur}, one recovers 
the ohmic part of
the current in eq.~\eqref{langevin} with the isospin conductivity
\begin{align}
\label{eq:quark isospin conductivity}
\sigma_q  = \frac{\mmWq}{3 \, \cw} \ ,
\end{align}
where 
\begin{align}
    \mmWq = \frac{\nc \nfam}{12} \gw^2 T^2 
    \label{sigq} 
\end{align}
is the quark contribution to the $ W $ boson screening mass.
So far, we have only accounted for quarks. 
Including all Standard Model fields, the conductivity turns out to be 
\cite{Bodeker:1998hm}
\begin{align}
     \sigma &= \frac{\mmW}{3 \, \cw}  
     \ ,
     \label{sigw}
\end{align}
which is identical to eq.~\eqref{eq:quark isospin conductivity},
except that $\mWq$ is replaced by the full screening mass $\mW$.
This remarkable similarity arises from the simplicity of the collision term in eq.~\eqref{eq:weak collision term},
which turns out to be the same for all Standard Model particles that carry weak isospin.
In general, quarks also
scatter with other isospin charged particles,
so that their collision terms will ``mix'' quark and other distribution functions, causing them to influence each other.
However, the term that mixes the distribution functions 
vanishes at next-to-leading log order when applied to an odd function of~$\vec v$,
so this mixing does not affect the $ l =1 $ sector of the collision term, 
which is odd under $ \vec v \to -\vec v $ \cite{Bodeker:1998hm}.
Hence, the distribution functions effectively decouple at the level of the long-wavelength limit of the Boltzmann equation given in eq.~\eqref{eq:lw boltzmann}.
As a result, generalizing eq.~\eqref{eq:quark isospin conductivity} to include all isospin-carrying particles amounts
to simply counting their isospin, color, and flavor degrees of freedom.

For our purposes, the absence of mixing terms also has the important consequence that
QCD corrections to the collision term only affect the quark distribution functions,
so that we do not have consider the out-of-equilibrium dynamics of the other 
Standard Model particles.

\section{The QCD collision term } 
\label{sec:qcd coll}

We now include the scattering of quarks
off quarks and gluons via QCD interactions. 
These do not mix quarks with other particles
that carry weak isospin. 
Therefore one can write the complete linearized
collision term as 
a sum of the weak-interaction term discussed in sec.~\ref{sec:vlasov} 
and the new strong-interaction term.
Like ref.~\cite{Arnold:2000dr},
we work at leading logarthmic order in the strong coupling $ \gs $.
We use the collision term in the form obtained 
in ref.~\cite{Ghiglieri:2018dib}, and 
also derive it using Dyson-Schwinger equations
in app.~\ref{eq:SD collision term}.

\begin{figure}
\centering
\begin{subfigure}[T]{.24\textwidth}
\includegraphics[width = \textwidth, clip, trim = 0 0 0 0]{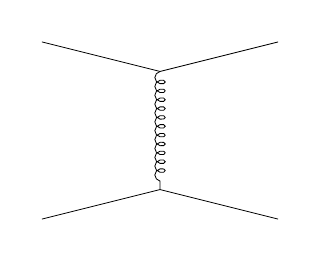}
\subcaption{\label{subfig:qq_qq_diagram} $qq \leftrightarrow qq$}
\end{subfigure}
\begin{subfigure}[T]{.24\textwidth}
\includegraphics[width = \textwidth, clip, trim = 0 0 0 0]{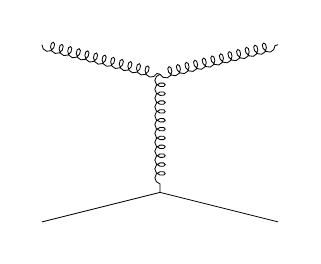}
\subcaption{\label{subfig:qg_qg_glex_diagram} $qg \leftrightarrow qg$}
\end{subfigure}
\begin{subfigure}[T]{.24\textwidth}
\includegraphics[width = \textwidth, clip, trim = 0 0 0 0]{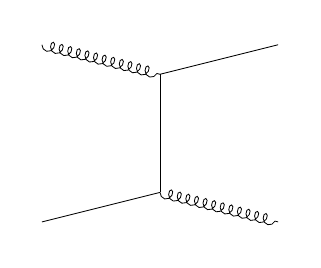}
\subcaption{\label{subfig:qg_qg_quex_diagram} $qg \leftrightarrow qg$}
\end{subfigure}
\begin{subfigure}[T]{.24\textwidth}
\includegraphics[width = \textwidth, clip, trim = 0 0 0 0]{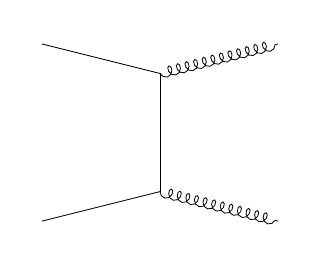}
\subcaption{\label{subfig:qq_gg_diagram} $qq \leftrightarrow gg$}
\end{subfigure}
\caption{\label{fig:feynman diagrams}%
Feynman diagrams contributing to the QCD collision term at leading-log order.
The logarithmic enhancement is the result of apparent IR divergencies
in the $ t $-channel contribution to $2\leftrightarrow2$ scattering 
processes with
soft gluon or quark exchanges.
Interference terms and $ s $-channel contributions
are IR safe and therefore only contribute at leading order.
}
\end{figure}

At leading logarithmic order,
only the $ 2\leftrightarrow2 $ processes with $t$-channel 
momentum exchange depicted in 
figure \ref{fig:feynman diagrams} contribute.
If the resulting collision term
were computed using tree-level propagators, 
the integral over the exchanged 4-momentum $ q $
would contain terms that
behave as $ \int d | \vec q | /| \vec q | $.
They are cut off by Bose-Einstein or Femi-Dirac distributions for 
large momenta
$ | \vec q | \gg T $,
but diverge logarithmically for small
$ | \vec q |  $.
However, medium effects significantly modify the propagation of the 
exchanged quarks and
gluons when $ | \vec q| $ is of order of the Debye scale $ \gs T $,
and provide a cutoff at even smaller momenta.
Due to the separation between hard ($ | \vec q | \sim T $) and 
soft ($ |\vec q | \sim \gs T $) scales,
one then obtains contributions that are enhanced by a factor
\begin{align}
\int\displaylimits_ { \gs T } ^ T  \frac{ d | \vec q |}{| \vec q |} 
   = \log ( \gs ^{ -1 } ) \ .
\end{align}
The leading-log approximations consists in keeping only these enhanced 
contributions.
Since $s$-channel exchanges and interference terms are infrared safe,
they do not contribute at leading logarithmic order.

Without loss of generality, we may consider a SU(2) electric field that it is diagonal in flavor space,
\begin{align}
\label{eq:E field ansatz}
   \vec E = \vec E ^ 3 \frac {\tau^3} 2 \ ,
\end{align}
where $\tau^3$ is the third Pauli matrix. 
Then the long-wavelength limit of the Boltzmann 
equation~\eqref{eq:lw boltzmann} for left-handed quarks of flavor $f$ takes 
the form
\begin{align}
\label{eq:ll quark boltzmann}
   \pm  \frac {\gw}2 \vec v \cdot \vec E ^ 3 \fF ' 
      =& - \cw \, \delta f_{fL} +
    \left [ \dot f _ {fL} \right ] _\text{strong} \ ,
\end{align} 
where the upper and lower sign applies to up- and down-type quarks, 
respectively.
We write the phase space densities as
\begin{align}
\label{eq:devan}
   f _ \alpha  = f _ \alpha  ^ {\rm eq } + f _ \alpha  ^ { \rm eq } 
   ( 1 \pm f _ \alpha ^ {   \rm eq } ) \eta  _ \alpha \ ,
\end{align}
where the index $ \alpha  $ denotes species and helicity. 
Then, one has
\begin{align}  
    \eta  _ { u L } = - \eta  _ { d L }, \quad  
   \eta  _ { q  L } = - \eta  _ { \bar q R } \ .
  \label{etaq} 
\end{align}
The right-handed quarks and left-handed antiquarks as well as the 
gluons remain in equilibrium, so that
$  \eta  _ { q R } = \eta _ {\bar q   L } = \eta  _ g = 0 $.
The linearized collision term due to $t$-channel gluon exchanges,
which mediate the quark-quark and quark-gluon scatterings shown in Figs.~\ref{subfig:qq_qq_diagram}
and \ref{subfig:qg_qg_glex_diagram},
can be written as \cite{Ghiglieri:2018dib}
\begin{align}
  \left [ \dot f _ \alpha   ( x,\vec p ) \right ] _ { \text{gl} }  
   = 
    \frac { \gs ^ 2 \log ( \gs ^ { -1 } )  } { 6 \pi  }
   \mmD T
  \partial _ { p ^ i } 
  \left \{
      \fF ( p _ 0 )  [ 1 - \fF ( p _ 0 ) ]
      \partial _ { p ^ i } \eta  _ \alpha  
  \right \} 
  + \mbox{ gain terms }  \ , 
   \label{collqq} 
\end{align} 
where $ \mD $ with
\begin{align} 
   \mmD &= \left( \frac{\nc}3 + \frac{\nfam}3 \right) \gs^2 T^2 \ 
   \label{mg} 
\end{align}
is the gluon Debye mass for $ 2\nfam$ quark flavors.
The gain terms vanish because they are linear operators acting on 
either $ \eta _  g = 0$
or the sum over quark flavors $ \sum _ f \eta  _ { f h } $, 
which vanishes due to  eq.~\eqref{etaq} for both helicities $ h=L,R $.

The analogous linearized collision term due to quark exchange,
which mediate the quark-gluon scatterings shown in 
Fig.~\ref{subfig:qg_qg_quex_diagram} and
the quark-antiquark annihilation shown in
Fig.~\ref{subfig:qq_gg_diagram}, is \cite{Ghiglieri:2018dib}   
\begin{align}
\label{eq:collqqgg}
\left [ \dot f _ { u L }   ( x,\vec p ) \right ] _ {\text{qu}} &=
- \frac{\gs^2 \log(\gs^{-1})}{6\pi p_0} \mmq
   [ 1+2 \fB(p_0) ] 
   \fF ( p _ 0 )  [ 1 - \fF ( p _ 0 ) ] \eta_\alpha 
+ \text{gain terms} \ ,
\end{align}
where
\begin{align}
\mmq &= \frac{\nc^2 - 1}{8 \nc} \gs^2 T^2
\end{align}
is the QCD contribution to the asymptotic quark mass~\cite{Arnold:2002ja}.%
\footnote{ We have neglected the electroweak contribution to the 
asymtpotic mass, assuming $ \gw\ll\gs $.}
Eq.~\eqref{eq:collqqgg} again contains no gain terms,
which vanish as in the electroweak case due to the rotational symmetries 
of the collision integral.

Like the linearized electroweak collision term,
the linearized QCD collision term commutes with rotations, so 
that $ \eta  _ \alpha  $ has the same angular dependence as the
force term.
For the left-handed up quarks we can thus write 
\begin{align}
    \eta  _ { u L } = 
   \frac {\gw} { 2 T } 
   \,\vec v \cdot \vec E ^ 3 \, \chi ( p _ 0 ) \ .
   \label{etachi}
\end{align} 
Inserting this into eqs.~\eqref{collqq} and \eqref{eq:collqqgg} gives 
\begin{align}
\label{qqchi}
\left [ \dot f _ { u L }   ( x,\vec p ) \right ] _ { \text{gl} } &=
\begin{multlined}[t]
    \frac {  \gs ^ 2 \log ( \gs ^ { -1 } ) } { 12 \pi  }  
    \mmD
   \gw 
   \, \vec v \cdot \vec E ^ 3\,
           \fF ( p _ 0 )  [ 1 - \fF ( p _ 0 ) ]
\\
  \times \left \{ 
      \chi '' + \left [ 
                    \frac 2 { p _ 0 } -  
                   \frac 1T
           [  1 - 2 \fF (p _ 0 ) ]  
                \right ] \chi '
                -  \frac 2 { p _ 0 ^ 2 } \chi 
   \right \} 
\end{multlined} 
\intertext{were primes denote derivatives with respect to $ p _ 0 $,
and}
\label{qqggchi}
\left [ \dot f _ { u L }   ( x,\vec p ) \right ] _ {\text{qu}} &=
- \frac{\gs^2 \log(\gs^{-1})}{12\pi}
\mmq \,
\gw 
\,\vec v \cdot \vec E ^ 3\,
\frac{1-2 \fB(p_0)}{T p_0}  \fF ( p _ 0 )  [ 1 - \fF ( p _ 0 ) ] \chi \ .
\end{align}
Finally, insertig the sum of these two
expressions into eq.~\eqref{eq:ll quark boltzmann},
one obtains 
a linear differential equation
\begin{align}
   1 = 
\cw \chi  
- \frac { \gs^2 \log ( \gs ^{ -1 } ) } { 6 \pi  }
   \Bigg\{ & 
     \mmD T
\left [  
      \chi '' + \left (
                    \frac 2 { p _ 0 } 
            - \frac 1T [  1 - 2 \fF ( p _ 0 ) ]  
                \right ) \chi '
                -  \frac 2 { p _ 0 ^ 2 } \chi 
   \right ]   
   \nonumber \\ 
   &
   - \mmq
   \frac { 1 + \fB ( p _ 0 ) } { 1 - \fF ( p _ 0 ) } 
   \frac { \chi  } { p ^ 0 } 
   \Bigg \} 
   \label{together} 
\end{align}
that determines the function $\chi(p_0)$.

\section{Results} 
\label{sec:numerics}

To obtain the quark contribution to the weak-isospin conductivity,
we have to solve eq.~\eqref{together} numerically.

\begin{table}
    \caption{Values for the parameter $\gamma $ entering 
         the $ l=1 $ eigenvalue of the electroweak collision term $ \cw $
         in eq.~\eqref{c1w} and
         the parameter $ a $ in the differential 
         equation~\eqref{eq:F equation}. 
         $\sigma_{\cancel {\rm QCD } }$ is the weak-isospin
         conductivity without QCD corrections,
         $\sigma_{\rm ana }$ is obtained using the analytic 
         approximation \eqref{eq:kapprox} for $\kappa$,
         and $\sigma_{\rm num}$ by solving eq.~\eqref{eq:F equation}  
         numerically.
      } 
  \begin{center}
\vspace{-16pt}
    \begin{tabular}{c|c|c|c|c|c|c} 
  $T$/GeV & $\gamma/T  $ &  $ \cw /T $ & $ a $ & $\sigma_{\rm \cancel{ \rm QCD } }/T  $ 
  & $\sigma_{\rm ana} / T$ & $\sigma_{\rm num } /T $ \\
      \hline
      130 & 0.127 & 0.327    & 1.521 & 0.773 & 0.728 & 0.728 \\
      200 & 0.126 &  0.325   & 1.602 & 0.773 & 0.729 & 0.730 \\
      $ 10^3 $  & 0.124 & 0.317 & 1.917 & 0.771 & 0.733 & 0.734 \\
      $ 10^8 $  & 0.107 & 0.271 & 4.609 & 0.763 & 0.742 & 0.745 \\
    \end{tabular}
    \label{t:gamma}
  \end{center}
\end{table}

In ref.~\cite{ Arnold:2003zc}, it was found that the complete leading 
order results for various transport coefficients are well approximated
by a next-to-leading log approximation that replaces 
\begin{align}
  \log(\gs^{-1}) &\to \log\left( \frac{\mu_\ast}{\mD} \right)  
  \qquad 
  \mbox{ with }  
  \qquad \mu_\ast \simeq 3 T
  \label{nll} 
\end{align}
in the leading-log result.
For our numerical results,
we adopt the prescription \eqref{nll}.
For convenience, we also define the dimensionless quantities
\begin{align}
   \label{diml} 
   a &= \frac{3 \pi \cw T}
    {g_s^2 \mmD \log\left( \mu_\ast/\mD \right)} 
   \ , &
   b &= \frac{m^2_\infty}{\mmD} = \frac16 
   \ , &
   x &= \frac{p_0}{T} 
   \ , &
   F(x) &= \frac{\cw \, \chi(p_0)}{a}
   \ .
\end{align}
Hence, the differential equation \eqref{together} for $ \chi $ turns into
\begin{align}
\label{eq:F equation}
1 &=
- \frac12 F^{\prime\prime}
- \left( \frac1{x} - \frac12 \frac{e^x - 1}{e^x + 1}  \right) F^\prime 
+ \left( a + \frac{b}{2 x} \frac{e^x + 1}{e^x - 1} + \frac1{x^2} \right) F \ ,
\end{align}
where the primes now denote derivatives with respect to $ x $.
The coefficient 
\begin{align}
 a \sim \frac { \gw ^ 2 \log ( \gw ^{ -1 } ) } 
  { \gs ^ 4 \log ( \gs ^{ -1 } ) }  
\end{align} 
parametrizes the relative size of the weak and strong collision terms.
For $ a \to { \infty  } $, one obtains the known result for the 
weak-isospin conductivity that neglects strong interactions altogether.
Taking $ a \to 0 $,
one instead obtains the weak-isospin conductivity in QCD.
The quark contribution to the weak isospin conductivity is
now given by
\begin{align}
   \sigma_q =  \frac{m_{W \text{qu}}^2}{3 \, \cw} \kappa 
   \ ,
\end{align}
where
\begin{align}
\label{eq:kappa def}
   \kappa = \frac{6 \, a}{\pi^2} 
   \int\displaylimits_0^\infty d x \, \frac{x^2 e^x F(x)}{(e^x + 1)^2} 
\end{align}
is normalized such that one has $\kappa \to 1$ for $a \to \infty$.

\subsection{ Boundary conditions}  
\label{s:dgl} 

To solve eq.~\eqref{eq:F equation} numerically,
one has to supply boundary conditions
at $x \to 0$ and $x \to \infty$.
Expanding eq.~\eqref{eq:F equation} for small $x$,
one obtains the approximate equation
\begin{align}
\label{eq:Fsmalleq}
1 &=
- \frac12 F^{\prime\prime}
- \frac1{x} F^\prime 
+ \frac{b + 1}{x^2} F \ ,
\end{align}
which has the general solution
\begin{align}\label{eq:smallxsol}
F(x) &= \frac{x^2}{b - 2} + k_1 \, x^{n_+} + k_2 \, x^{n_-} \ , &
n_\pm &= - \frac12 \pm \frac12 \sqrt{ 1 + 16 (b+1)^2 }
\end{align}
with integration constants $k_{1,2}$. 
The Pauli exclusion principle implies
that the quark distribution function $f_{fL}$ must remain finite as $x \to 0$, so that $k_2$ has to be set to zero.
Since the remaining terms in eq.~\eqref{eq:smallxsol} vanish for small $x$,
this fixes the boundary condition $F(x) \to 0$ as $x\to 0$.
For large $x$, one finds the approximate equation
\begin{align}
\label{eq:Flargeeq}
1 &=
- \frac12 F^{\prime\prime}
+ \frac12 F^\prime 
+ a F \ ,
\end{align}
which has the general solution
\begin{align}
F(x) &= \frac1{a} + k_3 \, e^{\lambda_+} + k_4 \, e^{\lambda_-} \ , &
\lambda_\pm &= \frac12 \pm \frac12 \sqrt{ 1 + 16 a^2}
\end{align}
with integration constants $k_{3,4}$.
Like before, the Pauli exclusion principle implies that $k_3$
has to be set to zero to ensure that $f_{fL}$ remains finite.
Therefore, the second boundary condition
is $F(x) \to 1/a $ for $x \to \infty$.

To obtain the numerical solution for the conductivity shown in 
fig.~\ref{fig:conductivity},
we have integrated \eqref{eq:F equation} 
from $x_\text{min} = 10^{-3}$ to $x_\text{max} = 30$
while enforcing boundary conditions $F(x_\text{min}) = 0$ and 
$F(x_\text{max}) = 1/a $.
We checked that varying $x_\text{min}$ in the range from $10^{-4}$ to $10^{-2}$ 
and $x_\text{max}$ from $20$ to $40$ barely affects the final result.

\subsection {Variational approach and analytic approximation} 

Solving the differential equation for $ F $ is equivalent to
solving a variational problem $ \delta  Q = 0 $
with a functional $ Q [ F ] $ containing terms
that are either linear and quadratic
in $ F $~\cite{Heiselberg:1994vy}.
\begin{figure}[tb]
\centering
\includegraphics[width = .55\textwidth, clip, trim = 15 0 5 0]{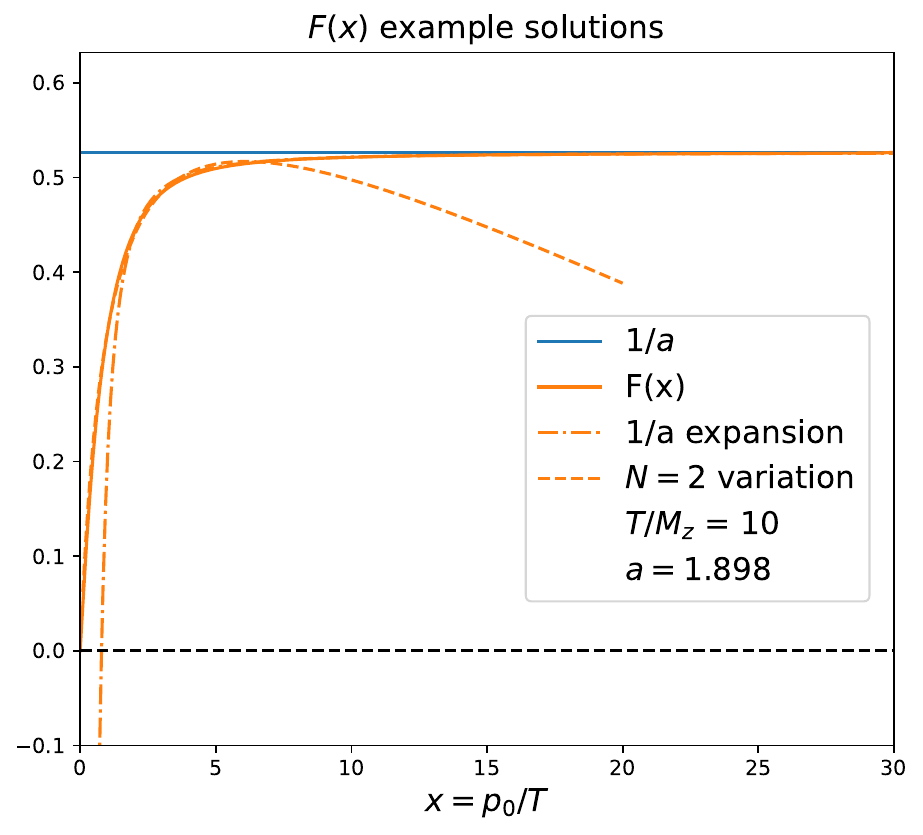}
\caption{\label{f:Fsol} 
   Solutions  for $F(x)$ at $ T/M _ Z =10$.
   The solid line shows  the results from direct numercal integration 
   of eq.~\eqref{eq:F equation},
   which is indistinguishable from the variational result with 
   $N = 10$ basis functions.
   The dash-dotted line shows the $1/a$ expansion in 
   eqs.~\eqref{eq:Fapprox}, and 
   the dashed line the result of the $ N = 2 $ variational ansatz.} 
\end{figure}
This is possible because the linearized collision term can be 
written as a functional derivative 
of a quadratic form \cite{Ghiglieri:2018dib}.
In our case,
we have to maximize the integral
\begin{align}
   Q [ F ]  =
   \int\limits _ 0 ^ { \infty  } 
   \frac { x ^ 2 e ^ x } { ( e ^ x + 1 ) ^ 2 } 
   \left \{ 
      F - \frac 14 F'{}^ 2
      - \left ( 
           \frac a2 + \frac b { 4x } 
           \frac { e ^ x + 1 } { e ^ x -1 } 
         + \frac 1 { 2 x ^ 2 } 
      \right ) 
      F ^ 2 
   \right \} \ .
   \label{Q} 
\end{align} 
We use the Ansatz
\begin{align}
   F ( x ) = \sum _ {  m =1 } ^ N c _ m \varphi  _ m ( x ) \ ,
\end{align}
with $N$ basis functions \cite{Arnold:2000dr}
\begin{align}
   \varphi  _ m ( x ) 
   =
   \frac { x ^ m } { ( 1 + x ) ^ { N -1 } }
   \qquad  ( m = 1, \ldots,  N) 
   \label{basis} 
   \ .
\end{align}
The maximum of $ Q $ is then found by 
solving $ \partial Q /\partial c _ m =0 $, which determines the~$ c _ m $.
Already with $ N =2 $ basis functions,
we find that the result for $ F( x )$
is a good approximation of the direct numerical solution
of eq. \eqref{eq:F equation} for not too large values of~$x$.
For example, when $a = 1.8$ and $1<x < 5$, it differs from the direct 
numerical solution  by less than 1.5\%.
Using the value of $c_1$ that follows from the variational approach 
and setting $c_2 \to 0$,
so that only the function $ \varphi _ 1 ( x ) = x/(1+x) $ from this 
basis contributes to $\kappa$, we find the analytic approximation
\begin{align}
\label{eq:kapprox}
   \kappa  = \alpha  \frac a { a + \beta  } \qquad 
   \mbox{ with } \qquad \alpha  = 0.9767 , \quad \beta  = 0.2107 \ .
\end{align} 
In the temperature range shown in Fig.~\ref{fig:conductivity},
this expression deviates from the numerical result by less than 1.1\%.
 
\subsection{Weak-isospin conductivity}
\label{s:con} 

\begin{figure}[tb]
\centering
\includegraphics[width = .48\textwidth, clip, trim = 5 0 460 0]{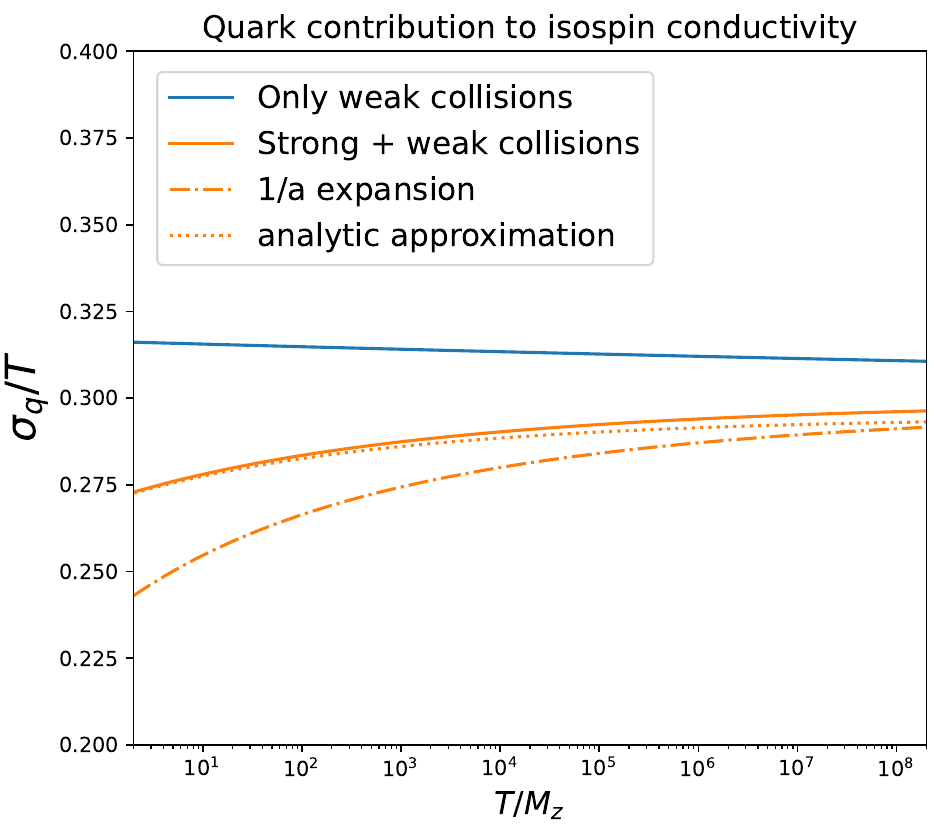}
\includegraphics[width = .48\textwidth, clip, trim = 465 0 0 0]{./figures/sigmaratio}
\caption{\label{fig:conductivity}%
Quark contribution to the isospin conductivity with and without the strong collision term.
The ratio $\kappa$ is defined in eq.~\eqref{eq:kappa def}.
Solid lines show results from direct numercial integration of 
eq.~\eqref{eq:F equation},
which are identical to the variational results with $N = 10$ basis functions.
Dash-dotted lines show results of the $1/a$ expansion given in eqs.~\eqref{eq:Fapprox} and \eqref{eq:1/a_kappa},
and dotted lines show the analytic approximation eq.~\eqref{eq:kapprox}.
The gauge couplings $\gs$ and $\gw$ are evaluated at
the renormalization scale $\rscale = \pi T$, see app.~\ref{sec:couplings} for details.}
\end{figure}

Figure \ref{fig:conductivity} depicts the quark contribution
to the weak-isospin conductivity for temperatures
above the electroweak crossover.
Overall, QCD interactions 
decrease the quark contribution to the conductivity by about 10 to 15\%
and are most relevant for low temperatures.
The full lines are obtained by direct numerical integration of
eq. \eqref{eq:F equation}.
This treats the electroweak and the QCD collision terms on equal footing, 
and corresponds to the power counting
\begin{align}
   \gw ^ 2 \log ( \gw ^{ -1 } ) \sim \gs ^ 4 \log ( \gs ^{ -1 } ) \ ,
   \label{pow} 
\end{align} 
or equivalently $ a \sim 1 $.
Alternatively, one may want to count $ \gs ^ 4 \ll \gw ^ 2 $
or $ a \gg 1 $, which is valid when
the QCD collision term is a small correction to the electroweak one.
Then, we can expand 
\begin{align}
\label{eq:Fapprox}
F(x) &= \frac1{a} - \frac1{a^2} \left( \frac1{x^2} + \frac{b}{2x} \frac{e^x + 1}{e^x - 1} \right) + \mathrm O(a^{-3}) \ .
\end{align}
The order $ a ^{ -2 } $ term gives a finite correction to
the integral in eq.~\eqref{eq:kappa def},%
\footnote{ 
Even higher orders in $ a ^{ -1 } $ would give rise to divergent integrals.
This is, however, 
no reason for concern, because there are other higher-order
contributions that we have not accounted for.}
so that one obtains
\begin{align}
\label{eq:1/a_kappa}
\kappa &= 1 - \left( \frac1{\pi^2} + \frac{b}8 \right) \frac3a + \mathrm O(a^{-2}) \ .
\end{align}
Treating the QCD collision term as a perturbation in this way,
one obtains the dash-dotted lines.
The result deviates from the full result by 
another 5 to 10 \%. 
Thus, it appears more sensible to fully include the QCD interactions.
For this, eq.~\eqref{eq:kapprox} (drawn in dotted lines) 
provides a simple analytic approximation
that works astonishingly well,
deviating form the numerical result by less than 1.1\%.

\section{Summary and conclusions}
\label{s:concl}

In the high-temperature phase of the Standard Model, when the 
expectation value of the Higgs field vanishes,
the dynamics of the weak SU(2) gauge fields is non-perturbative. 
It determines the so-called hot sphaleron rate, which in turn governs
the rate for baryon number violation.
The gauge field dynamics is overdamped, and the damping coefficient is 
the weak isospin conductivity.
We have computed QCD corrections to
this conductivity at leading lograthmic order,
and find that they reduce the value of the conductivity
by 6\% near the electroweak crossover.
As shown in table~\ref{t:gamma}, the size of the corrections decreases
at higher temperatures.
The corrected value can directly be used to update of
the hot sphaleron rate.
Including QCD corrections, the complete isospin conductivity is given by
\begin{align}
   \sigma = \frac 1 { 3 \cw } 
   \Big [ \mmW - ( 1 - \kappa  ) \mmWq
   \Big ]
   \ ,
   \label{sigma} 
\end{align}
where $\kappa$ parametrizes the impact of strong interactions.
A simple yet accurate analytic approximation for $ \kappa  $ is given
by eq.~\eqref{eq:kapprox}.

Our results show that QCD interactions give rise to a relatively small correction to
the weak-isospin conductivity, but also that the correction is 
big enough that one should  treat the QCD interactions 
on the same level as the weak interactions when solving the
Boltzmann equation.

We have worked at leading-log order both in the weak and in the
strong coupling. 
For the collision term this involved incorporating only 
the contributions of 
order $\gw ^ 2 \log(\gw ^{ -1  })  $ and $ \gs ^ 4 \log ( \gs ^{ -1 })  $.
Going beyond this approximation should  be relatively straightforward
for the QCD contribution. 
A first step would be to include the non-logarithmic leading order 
contributions
from $ 2 \leftrightarrow 2 $ scatterings and near-collinear in-medium 
emissions, which have been computed for QCD transport 
coeffcients~\cite{Arnold:2003zc}.
Additionally, one could also incorporate order $ \gs $ 
corrections~\cite{Caron-Huot:2008zna},  
which were computed for QCD transport coefficients 
in Ref.~\cite{ Ghiglieri:2018dib}. 
It is less clear how to go beyond the leading-log approximation
in the purely electroweak sector. 
While it would be possible to include the relevant corrections to the collision term,%
\footnote{ 
   For example,
   one could incorporate order $ \gs ^ 2 $ thermal quark masses
   in the weak collision term, which would result in 
   $ \gw ^ 2 \gs ^ 2 $ contributions. 
   However, there are 
   additional scattering process that contribute at the same order,
   and that would need to be included on the same footing to achieve a fully 
   consistent result.
   It is also important to note that these contributions are of higher order   
   than the order $ \gw ^ 2 $ 
   part of the  
   collision term without logarithmic enhancement that we have neglected. 
}
one could then no longer neglect the Liouville term on the left-hand side
of the Boltzmann equation~\eqref{eq:lw boltzmann} \cite{Bodeker:2000da,Bodeker:2002gy}, 
which significantly complicates the computation.
This would also generate additional operators on the right-hand side of 
Eq.~\eqref{langevin},
so that the impact of higher momentum
modes on the evolution of the magnetic-scale 
gauge fields 
is no longer characterized by a single conductivity.

\section*{Acknowledgments}

We thank G.D.~Moore for insightful comments.
DB and PK acknowledge support by the Deutsche Forschungsgemeinschaft 
(DFG, German Research Foundation) through the CRC-TR 211 
'Strong-interaction matter under extreme conditions'– 
project number 315477589 – TRR 211.
PK was also supported by the Swiss National Science Foundation 
(SNSF) under grant P500PT-217885.

\appendix

\section{Coupling constants}
\label{sec:couplings}

To numerically evaluate the quark contribution to the isospin conductivity,
one has to specify values for the coupling constants $\gs$ and $\gw$, which entails choosing a renormalization scale $\rscale$.
At one loop, the couplings are%
\footnote{See, e.g.,  \cite{Gunion:1989we}.} 
\begin{align}
\frac1{\gw^2} &= \frac1{\gw^2(\mz)} 
    + \frac{\betaw}{8\pi^2} \log \left( \frac{\rscale}{\mz} \right)
   \label{gw} 
   \ , \\
\frac1{\gs^2} &=  \frac1{\big[\gs^{(5)} (\mz)\big]^2} 
   + \frac1{12\pi^2} \log \left( \frac{m_t}{\mz} \right) 
   + \frac{\betas}{8\pi^2} \log \left( \frac{\rscale}{\mz} \right) 
   \label{gs} 
   \,
\end{align}
where $\mz = \unit[91.1880\pm0.0020]{GeV}$ is the $ Z $ Boson mass
and $\gw(\mz) = 0.65163 \pm 0.00008$ the weak gauge coupling 
at $\rscale = \mz$ in the modified minimal subtraction ($\overline{\text{MS}}$) 
scheme \cite{ParticleDataGroup:2024cfk}.
Furthermore,
$\gs^{(5)}(\mz) = 1.218 \pm 0.005$ the strong gauge coupling 
at zero temperature and at $\rscale = \mz$, i.e., below the top quark mass
$m_t = \unit[162.69\pm0.006]{GeV}$, where only 5 quark flavors are active.%
\footnote{
   Ref.~\cite{ParticleDataGroup:2024cfk}
   reports a $1.7 \sigma$ discrepancy between the pole mass from direct 
   measurements at the LHC $m_t^\text{pole} = \unit[172.52\pm0.44]{GeV}$
   and the mass from electroweak precision data 
   $m_t^\text{pole} = \unit[176.12\pm1.9]{GeV}$.
   To compute $\gs$, we use the $\overline{\text{MS}}$ mass
   $m_t = \unit[162.69\pm0.006]{GeV}$ that one obtains by converting 
   the direct measurement of the pole mass \cite{ParticleDataGroup:2024cfk}.
   We checked that using the pole-mass from electroweak precision data instead
   does not affect the first three significant digits of $\sigma /T$ 
   shown in tab. \ref{t:gamma}.
   }
Considering temperatures above the electroweak crossover, we use the
value of the strong coupling with $ 2 \nfam =6 $ active quark flavors,
which differs from the value with 5 flavors by the 
matching term proportional to $\log(m_t / \mz)$.
The beta function coefficients are 
\begin{align}
\betas &= \frac{11}3 \nc - \frac43 \nfam \ , &
\betaw &= \frac{22}3 - \frac16 - \frac43 \nfam \ .
\end{align}
Since the quark contribution to the weak-isospin current \ref{qcur} is dominated by hard momenta $p \sim \pi T$,
we choose a renormalization scale $\rscale = \pi T$.
This results in temperature dependent values of the coupling constants $\gs$ and $\gw$,
which is in fact the dominant source of the temperature dependence for the isospin conductivity shown in fig.~\ref{fig:conductivity}.

\section{Vlasov-Boltzmann from Schwinger-Dyson equations}  
\label{sec:vlasov-boltzman derivation}

In this appendix, we use a 2PI effective approach to derive Vlasov-Boltzmann 
equations for the left-handed quark distribution functions $f_{fL}$
in the presence of a soft and classical non-Abelian gauge field $A_\mu$
from first principles of non-equilibrium quantum field theory.
The derivation of Boltzmann equations via the 2PI approach is well understood
(see e.g. \cite{Arnold:1998cy,Prokopec:2003pj,Prokopec:2004ic,Berges:2004yj,Drewes:2016gmt,Garbrecht:2018mrp}),
but it is comparatively less established how to modify it to account for the 
classical gauge field while ensuring gauge covariance.
That being said, gauge covariant transport equations for non-Abelian gauge theories were previously derived in \cite{Elze:1989ba} and \cite{Blaizot:1993be},
and used in \cite{Blaizot:1993be} to derive Vlasov equations that are valid at leading order in a hard thermal loop expansion.

We follow the same overall approach as in \cite{Elze:1989ba},
and consider the time-evolution of the two point functions
\begin{align}
\label{eq:covpropdef}
i \hat S^{ab} (u,v) &=
W (x, u) \, \langle \mathcal T_C \big\{ q^{a}_{fL} (u) \otimes \overline q^{b}_{fL} (v) \big\} \rangle \, W (v ,x) \ , &
x &= \frac{u+v}2 \ ,
\end{align}
where the $q_{fL}$ are left-handed quark fields of flavor $f$, $\otimes$ is a tensor product in Dirac, isospin, and color space,
and $\langle \cdot \rangle$ a quantum-statistical average.
The time path-ordering operator $\mathcal T_C$ is defined such that fields are evaluated in sequence along the closed time-path $C$,
which begins at some initial time $t_i$ in the far past, $t_i \to - \infty$,
and goes forward until reaches some final time $t_f$ in the far future, $t_f \to + \infty$, where it turns around and returns back to $t_i$.
Therefore, the closed time path has two separate branches,
and the indices $a,b = \pm 1$ denote that the corresponding quark field is to be evaluated along either the forward (``$+$'') or backward (``$-$'') branch.
If both fields are evaluated along the same branch, $\mathcal T_C$ reduces to the standard (anti-)time ordering operators familiar form zero-temperature QFT.
Finally, the Wilson lines
\begin{align}
\label{eq:wilsondef}
W \big( x + \frac{r}2, x \big) &= \mathcal P \exp\left( i \gw \int\displaylimits_0^1 \text{d} t \, \frac{r^\mu}2 A_\mu \big( x + \frac{r}2 t \big) \right) \ ,
\end{align}
where $A_\mu$ is a soft and classical 
gauge field and $\mathcal P$ is a path-ordering operator,
ensure that $i \hat S(u,v)$ transforms
under the adjoint representation at $x$.
Here and in the following, we suppress isospin and color indices for the sake of readability,
but it should be remembered that $A_\mu$, $W$, and $i \hat S^{ab}$ are matrices in isospin and color space.
On the other hand, there is no flavor mixing because we neglect Yukawa interactions and
work in the symmetric phase of the Standard Model.

We primarily focus on the Wightman functions $\hat S^{>} = \hat S^{-+}$ and $\hat S^<= \hat S^{+-}$,
which encode information about the quark distribution functions
\begin{align}
\label{eq:distribution functions definition}
f_{fL} (\bm p)
&\equiv - \int \frac{\text{d} p_0}{2 \pi} \Theta(p_0) \ \text{tr}_D \left[ \gamma^0 i S^< \right] \ , &
f_{\overline fR} (- \bm p)
&\equiv \int \frac{\text{d} p_0}{2 \pi} \Theta(- p_0) \ \text{tr}_D \left[ \gamma^0 i S^> \right] \ ,
\end{align}
where $f_{\overline f R}$ is the distribution function of a right-handed anti-quark of flavor $f$,
and 
\begin{align}
i S^{ab} (x,p) &= \int \text{d}^4 r \, e^{i n r} \,
i \hat S^{ab} \big( x + \frac{r}2, x - \frac{r}2 \big) 
\end{align}
are the Wightman functions in Wigner-space.
Since the quarks are massless, we may parametrize the propagators as
\begin{align}
i S^{ab} &= P_L \gamma^\mu P_R i S^{ab}_\mu \ .
\end{align}
We also need the spectral, advanced and retarded combinations
\begin{align}\label{eq:spectral def}
S^{A} &= \frac1{2 i} \left( S^a - S^r \right) \ , &
S^{a,r} &= S^{++} - S^{>,<} = S^{<,>} - S^{--} \ .
\end{align}

To start, let us neglect the classical field $A_\mu$, 
so that the Wilson lines are trivial.
While it is well known how to derive Boltzmann equations in this limit, 
we provide a short outine of the approach to better set up the subsequent inclusion of the background field $A_\mu$.
At leading order in a gradient expansion and to any loop order, it can be shown
that the Wightman functions obey the kinetic equation \cite{Arnold:1998cy,Prokopec:2003pj,Drewes:2016gmt}
\begin{equation}
\label{eq:no wilson kinetic equation}
i \partial^\mu \, i S_\mu^{>,<} = \Sigma_\mu^r \, i S^{>,< \, \mu} + i \Sigma_\mu^{>,<} \, S^{a \, \mu} - i S_\mu^{>,<} \, \Sigma^{a \, \mu} - S_\mu^r \, i \Sigma^{>,< \, \mu} \ ,
\end{equation}
where the self-energies in Wigner space
\begin{align}\label{eq:quselfdef}
i \Sigma_\mu^{ab} (x,p) &\equiv \frac12 \int \text{d}^4 r \, e^{i p r} \, \text{tr}_D \left[ P_L \gamma_\mu i \hat \Sigma^{ab} \big( x + \frac{r}2, x - \frac{r}2 \big) \right]
\end{align}
and position space
\begin{align}
i \hat \Sigma^{ab} (u,v) &\equiv ab \frac{\delta \, i \mathrm \Gamma_2}{\delta \, i S^{ba} (v,u)}
\end{align}
are consistently defined as variations of the interaction part
of the 2PI effective action $\mathrm \Gamma_2$ \cite{Cornwall:1974vz,Prokopec:2003pj,Berges:2004yj}.
The advanced, retarded, and spectral selfenergies are defined analogously to eq. \eqref{eq:spectral def}.
At linear order in deviations from equilibrium, the projected self-energies $i \Sigma^{ab}_\mu$ and propagators $i S^{ab}_\mu$ commute, yielding
\begin{align}
\label{eq:quantum kinetic equation}
\frac12 \partial^\mu \, i S_\mu^{>,<} = - \Sigma_\mu^{\mathrm A} \, i S^{>,< \, \mu} + i \Sigma_\mu^{>,<} \, S^{\mathrm A \, \mu} \ .
\end{align}
To make contact with kinetic theory, we evaluate this equation at leading order in a perturbative loop expansion,
inserting the tree-level Kadanoff-Baym Ansatz
\begin{subequations}\label{eq:qutprop}
\begin{align}
[i S_\mu^>]_\text{tree} &= p_\mu \, 2 \pi \, \text{sign} (p_0) \delta (p^2) \left[ (1 - f_{fL} (\bm p)) \Theta(p_0) + f_{\overline fR} (-\bm p) \Theta(-p_0) \right] \ , \\
[i S_\mu^<]_\text{tree} &= p_\mu \, 2 \pi \, \text{sign} (p_0) \delta (p^2) \left[ - f_{fL} (\bm p) \Theta(p_0) - (1 - f_{\overline fR} (-\bm p)) \Theta(-p_0) \right] \ .
\end{align}\end{subequations}
Hence, multiplying both sides of eq.~\eqref{eq:quantum kinetic equation} by $\Theta(p_0)$, integrating over $p_0$,
adding up the $>$ and $<$ cases, and linearizing in the deviation from equilibrium, one obtains
\begin{align}
\label{eq:quark boltzmann trivial}
v^\mu \, \partial_\mu f_{fL} (\bm p) &=
\left[ \dot f_{fL} \right]_\text{coll} \ , 
\end{align}
where  $v^\mu \equiv (1, \nicefrac{p^i}{p_0})$ and the collision term is given as
\begin{align}
\label{eq:collision term definition}
\left[ \dot f_{fL} \right]_\text{coll} &= 
v^\mu \left[ -2 f_{fL} (\bm p) \Sigma_{\mu \, \text{eq}}^{\mathrm A} + \delta \Sigma_\mu^{\mathrm A} ( 1 - 2 \fF(p_0) ) - i \delta \Sigma_\mu^+ \right] \ .
\end{align}

It remains to include the classical field $A_\mu$.
Since the leading collision term on the right-hand side of eq.~\eqref{eq:quark boltzmann trivial} is already linear in the deviation from equilibrium,
we only need to modify the Liouville operator $v^\mu \partial_\mu f_{fL}$ on the left-hand side.
This can be done by considering the time-evolution of the Wightman functions in a free theory,
where they obey the Dirac equation
\begin{align}
i \slashed D (u) \left[  W(u, x) \, i \hat S^{>,<} (u,v) \, W(x,v) \right] &= 0 \ , &
i D_\mu (u) &= i \partial_\mu^u + g A_\mu (u) \ .
\end{align}
Keeping in mind that $x = \nicefrac{(u+v)}2$, the derivatives of the Wilson lines are \cite{Elze:1989ba}
\begin{subequations}\begin{align}
i \partial_\mu^u \, W(u, x) &=
- g A_\mu (u) W(u,x) + \frac{g}2 W(u,x) A_\mu (x) 
- G_1(u,x)
\ , \\
i \partial_\mu^u \, W(x,v) &=
- \frac{g}2 A_\mu (x) W(x,v)
- G_0(x,v) \ ,
\end{align}\end{subequations}
where
\begin{align}
G_n (u,x) &= 
g \int\displaylimits_0^1 \text{d} t \, \frac{n+t}2 W(u, u_t) \, \dot u_t^\alpha F_{\mu \alpha} (u_t) \, W (u_t, x) \ , &
u_t &= x + (u-x)t \ .
\end{align}
Hence, defining $r = u-v$, one obtains
\begin{multline}
\left( i \slashed \partial_r + \frac{i}2 \slashed D_\text{adj} (x) \right) i \hat S^{>,<} (u,v) \\
- W(x,u) \, G_1(u,x) \, i \hat S^{>,<} (u,v) - i \hat S^{>,<} (u,v) \, G_0(x,v) \, W(v,x) = 0 \ ,
\end{multline}
where
\begin{align}
i D_{\mu \, \text{adj}} (x) \, i \hat S^{>,<} &\equiv i \partial^x_\mu \, i \hat S^{>,<} + g \big[ A_\mu (x), i \hat S^{>,<} \big] 
\end{align}
is now the covariant derivative in the adjoint representation.
Since the gauge field is soft, derivatives acting on the field strength tensor $F_{\mu\alpha} (u_t)$
scale as $D_t F_{\mu\alpha} \sim g^4 \ln(g^{-1}) F_{\mu\alpha}$ and $D_i F_{\mu\alpha} \sim g^2 F_{\mu\alpha}$.
Working at leading order in the coupling,
we can therefore neglect these gradients and replace $F_{\mu\alpha} (u_t) \to F_{\mu\alpha}(x)$ in $G_n$.
For the same reason, we may also neglect the gauge fields in the remaining Wilson lines, which gives
\begin{multline}
\left( i \slashed \partial^r + \frac{i}2 \slashed D_\text{adj} (x) \right) i \hat S^{>,<} (u,v) \\
- \frac{3g}8 \, r^\alpha F_{\mu\alpha} (x) \, \gamma^\mu i \hat S^{>,<} (u,v) - \frac{g}8 \, \gamma^\mu i \hat S^{>,<} (u,v) \, r^\alpha F_{\mu\alpha} (x) = 0 \ .
\end{multline}
To proceed, we transform to Wigner space, consider the imaginary part of the resulting equation,
and take a Dirac trace, which yields the kinetic equation
\begin{align}
\label{eq:quantum vlasov equation}
\frac12 i D_\text{adj}^\mu (x) \, i S_\mu^{>,<} (x,p) + i \frac{g}4 \, \big\{ F^{\mu\alpha} (x) , \partial^p_\alpha \, i S^{>,<}_\mu (x,p) \big\} &= 0 \ .
\end{align}
This is the equivalent of eq.~\eqref{eq:quantum kinetic equation}
for a free quark in the presence of a soft background field.
Proceeding as we have before for eq.~\eqref{eq:no wilson kinetic equation}  (i.e. inserting the tree-level expressions \eqref{eq:qutprop}
into eq.~\eqref{eq:quantum vlasov equation},
multiplying by $\Theta (p_0)$,
integrating over $p_0$,
adding up the $>$ and $<$ cases,
and finally linearizing in the deviation from equilibrium),
one recovers the Vlasov equation 
\begin{align}
v_\mu D_\text{adj}^\mu (x) \, f_{fL} (\bm p)
+ g E_i (x) v^i \fF^\prime (p_0) &= 0  \ , &
E_i &= F_{0i} \ . &
\end{align}
Comparing this result with the Boltzmann equation \eqref{eq:quark boltzmann trivial},
we see that the soft classical field can be included via the replacement
\begin{align}
v^\mu \partial_\mu \, f_{fL} (\bm p) &\to
v_\mu D_\text{adj}^\mu (x) \, f_{fL} (\bm p)
+ g E_i (x) v^i \fF^\prime (p_0) \ .
\end{align}
Indeed, keeping track of both the classical 
field and the interacting part of the 2PI effective action,
which generates the collision term, and repeating the same derivation,
one obtains the linearized non-Abelian Vlasov-Boltzmann equation
\begin{align}
\label{eq:full vlasov boltzmann}
v_\mu D_\text{adj}^\mu (x) \, f_{fL} (\bm p)
+ g E_i (x) v^i \fF^\prime (p_0) &= \left[ \dot f_{fL} \right]_\text{coll} \ .
\end{align}
Following the discussion in sec.~\ref{sec:vlasov}, we further neglect the $v_\mu D^\mu f_{fL}$ term
when computing QCD corrections to the isospin conductivity at leading-logarithmic accuracy.
Hence, inserting the expression for $E_i$ in eq.~\eqref{eq:E field ansatz} into eq.~\eqref{eq:full vlasov boltzmann},
one recovers eq.~\eqref{eq:ll quark boltzmann}.

\section{QCD collision term from 2PI effective action}
\label{eq:SD collision term}

In this appendix,
we compute the QCD collision term defined 
in eq.~\eqref{eq:collision term definition}.
\begin{figure}
\centering
\begin{subfigure}[c]{.28\textwidth}
\includegraphics[width = \textwidth, clip, trim = 0 0 0 0]{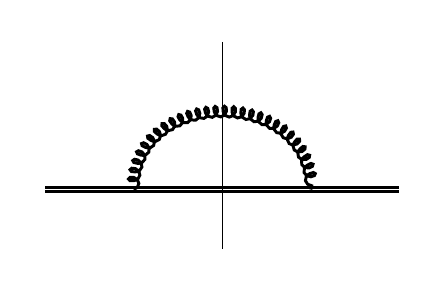}
\caption{\label{fig:1lresdiag}}
\end{subfigure}
\begin{minipage}[c]{.05\textwidth}
\Large
$\supset$
\end{minipage}
\begin{subfigure}[c]{.28\textwidth}
\includegraphics[width = \textwidth, clip, trim = 0 0 0 0]{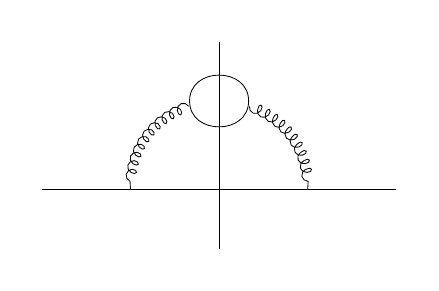}
\caption{\label{fig:glexdiag}}
\end{subfigure}
\begin{minipage}[c]{.05\textwidth}
\Large
$+$
\end{minipage}
\begin{subfigure}[c]{.28\textwidth}
\includegraphics[width = \textwidth, clip, trim = 0 0 0 0]{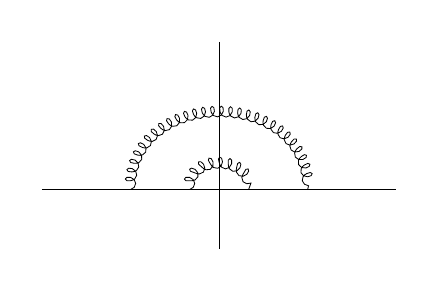}
\caption{\label{fig:quexdiag}}
\end{subfigure}
\caption{\label{fig:1lqcddiag}
Leading logarithmic contributions to the quark selfenergy.
The cut lines represent Wightman functions while uncut lines represent (anti-)time ordered propagators.
The thick double lines on the left-hand side denote exact propagators, which can be expanded perturbatively
to obtain the two diagrams on the right-hand side, thus yielding the relevant logarithmically enhanced contributions.
}
\end{figure}
To do so, one has to supply an apropriate expression for the self-energy \eqref{eq:quselfdef}.
At leading logarithmic order, it is sufficient to consider the resummed one-loop diagram shown in Fig.~\ref{fig:1lresdiag},
\begin{align}
\nonumber
v^\mu i \Sigma_{\mu}^{>,<}
&\equiv \frac1{2 p_0} \text{tr}_D\left[ P_L \slashed p \, i \Sigma^{>,<} (p) \right]
\\\label{eq:quark self energy}
&= - \frac{4 \gs^2}3 \int \frac{\text{d}^4 k}{(2\pi)^4} \frac{\left(2 p^\nu g^{\mu\rho} - p^\mu g^{\nu\rho} \right)}{p_0} \, i S^{>,<}_{\mu} (k) \, i \Delta^{>,<}_{\nu\rho} (p-k) \ ,
\end{align}
where $i S^{>,<} (k)$ and $i \Delta^{>,<}_{\nu\rho}$ are exact quark and gluon Wightman functions.
These functions can be expanded perturbatively to obtain the two unresummed two-loop diagrams shown in Figs.~\ref{fig:glexdiag} and~\ref{fig:quexdiag}.
Using the optical theorem, it can be shown that these diagrams encode the relevant logarithmically enhanced t-channel contributions
from $2\leftrightarrow2$ scattering processes shown in Fig.~\ref{fig:feynman diagrams}.

To extract these contributions, one has to insert appropriate tree-level and one-loop expressions for the Wightman functions into eq.~\eqref{eq:quark self energy}
while neglecting IR safe terms that do not contribute to the leading-log result.
Since gluons carry no isospin charge, their propagator does not depart from thermal equilibrium.
This greatly simplifies the computation, as one can use well-known thermal equilibrium expressions.

\subsubsection*{Gluon exchanges}

First, we consider the gluon exchange contribution depicted in Fig.~\ref{fig:glexdiag}.
This diagram encodes the same physics as the $qq \leftrightarrow qq$ and $qg \leftrightarrow qg$ diagrams depicted in Figs.~\ref{subfig:qq_qq_diagram} and \ref{subfig:qg_qg_glex_diagram}.
To extract this contribution, we replace
\begin{align}
i S_\mu^{>,<} &\to [i S_\mu^{>,<}]_\text{tree} \ , &
i \Delta^{>,<}_{G \, \nu\rho} &\to [i \Delta^{>,<}_{G \, \nu\rho}]_\text{one-loop} \ ,
\end{align}
where the tree-level quark propagator is the same as in eq.~\eqref{eq:qutprop}.
The thermal equilibrium gluon propagator
\begin{align}\label{eq:gleqprop}
i \Delta^<_{G \, \nu\rho} (q) &= 2 \fB(q_0) \, \rho_{\mu\nu} (q) \ , &
i \Delta^>_{G \, \nu\rho} (q)  &= 2 (1 + \fB(q_0)) \, \rho_{\mu\nu} (q) \ ,
\end{align}
is fully determined by the spectral function $\rho_{\mu\nu}$.
In a general covariant gauge, it is given as \cite{Braaten:1989mz,Gorda:2023zwy}
\begin{align}\label{eq:gl1lspec}
[ \rho_{\mu\nu} (q) ]_\text{one-loop} &=
\frac{- P^T_{\mu\nu} \mathrm \Pi^T - P^L_{\mu\nu} \mathrm \Pi^L}{q^4}
- \alpha \frac{q_\mu q_\nu}{q^2} \pi \, \text{sign} (q_0) \, \delta(q^2) \ ,
\end{align}
where $\alpha$ is a gauge-fixing parameter, $\mathrm \Pi^{L/T}$ are transverse and longitudinal spectral self-energies,
while
\begin{align}
P^T_{\nu\rho} &= \delta_\nu^i \delta_{\rho}^j \left( g_{ij} + \frac{g_{00} q_i q_j}{\bm q^2} \right) \ , &
P^L_{\nu\rho} &= - \frac{(q^2 g_{0\nu} - q_0 q_\nu) (q^2 g_{0\rho} - q_0 q_{\rho})}{q^2 \bm q^2}
\end{align}
are the corresponding projection operators.
Focussing on t-channel ($q^2 < 0$) contributions and dropping terms that are IR safe in the $|\bm q| \to 0$ limit,
we replace the full self-energies by their HTL expressions \cite{Braaten:1989mz}
\begin{align}
\label{eq:HTL self}
\mathrm\Pi^L &\overset{\text{HTL}}{=} - 2 \mathrm \Pi^T \overset{\text{HTL}}{=} \frac{\pi}2 \mmD \frac{q_0}{|\bm q|} \left( 1 - \frac{q_0^2}{\bm q^2} \right) \ .
\end{align}
Inserting the one-loop expressions~\eqref{eq:gleqprop} and \eqref{eq:gl1lspec} for the gluon propagator
and the tree-level quark propagator~\eqref{eq:qutprop} into the eq.~\eqref{eq:quark self energy},
one obtains a collision term that sums all contributions from $2\leftrightarrow2$
and kinematically allowed $1 \leftrightarrow 3 $ processes mediated by a gluon exchange while neglecting interference terms.
To isolate only t-channel exchanges, we require that the energy $k_0$ of the quark propagator 
in eq.~\eqref{eq:quark self energy} is positive.
Hence, using the Ansatz \eqref{eq:devan} to parametrize the deviation from equilibrium,
and dropping IR safe contributions,
one obtains the collision term
\begin{align}\nonumber
\left [ \dot f_{fL} (\bm p) \right ] _ \text{gl}
&= \frac{\mmD \gs^2}{4 \pi^2} \int \frac{\text{d}^3 \bm q}{| \bm q |^5} (1 - \fF(p_0)) \fF(p_0 - q_0) q_0 \fB(q_0) \left[ \eta_{fL} (\bm p - \bm q) - \eta_{fL} (\bm p) \right]
\\\label{eq:collprecutoff}
&= \frac{\mmD \gs^2 T}{6 \pi} \delta^{ij} \partial_{p_i} \left[ (1 - \fF(p_0)) \fF(p_0)  \partial_{p^j} \, \eta_{fL}(\bm p) \right] \int \frac{\text{d} | \bm q |}{| \bm q |} + \text{ IR safe} \ ,
\end{align}
where
\begin{align}
q_0 &\equiv | \bm p | - | \bm p - \bm q | = 
(\hat p \hat q) \frac{|\bm q|}{p_0} 
+ O( \nicefrac{|\bm q|^2}{p_0^2} )
\end{align}
is the energy of the exchanged gluon.
Note that this expression is manifestly independent of the gauge fixing parameter $\alpha$.
To finally extract the leading-logarithmic contribution, we follow e.g. \cite{Bodeker:1998hm,Arnold:2000dr} and impose the IR and UV cutoff
\begin{align}\label{eq:llcutoff}
\int \frac{\text{d} | \bm q |}{| \bm q|} &\to \int\displaylimits_{\gs T}^{T} \frac{\text{d} | \bm q |}{| \bm q|} = \log (\gs^{-1} ) \ .
\end{align}
Hence, inserting this replacement into eq.~\eqref{eq:collprecutoff},
we recover eq.~\eqref{collqq} for the gluon exchange contribution to the QCD collision term,
which is also given e.g. in \cite{Ghiglieri:2018dib}.

\subsubsection*{Quark exchanges}

Next, we consider the quark exchange contribution depicted in Fig.~\ref{fig:quexdiag}.
Up to interference terms, which are captured by separate two-loop diagrams, this diagram encodes the same physics as the $qg \leftrightarrow qg$
and $qq \leftrightarrow gg$ diagrams depicted in Figs.~\ref{subfig:qg_qg_quex_diagram} and~\ref{subfig:qq_gg_diagram}.
To extract this contribution, we replace
\begin{align}
i S_\mu^{>,<} &\to [i S_\mu^{>,<}]_\text{one-loop} \ , &
i \Delta^{>,<}_{\nu\rho} &\to [i \Delta^{>,<}_{\nu\rho}]_\text{tree} \ , 
\end{align}
where the tree-level gluon propagator is the same as in eq.~\eqref{eq:gleqprop},
except with the one loop spectral function replaced by the well-known tree-level expression
\begin{align}\label{eq:gleqspectree}
[ \rho_{\mu\nu} (q) ]_\text{tree} &= \left( - g_{\mu\nu} +  (1- \alpha) \frac{q_\mu q_\nu}{q^2} \right) \pi \, \text{sign} (q_0) \, \delta(q^2) \ .
\end{align}
The one-loop quark propagator
\begin{align}\label{eq:1lquprop}
[i S_\mu^{>,<} (q) ]_\text{one-loop}
&= \frac{2 q^\mu q^\sigma - q^2 g^{\mu\sigma}}{q^4} [ i \Sigma_{\sigma}^{>,<} (q) ]_\text{one-loop} 
\end{align}
is defined recursively in terms of the strict one-loop contribution to the quark self-energy \eqref{eq:quark self energy},
which is given by replacing both the quark propagator and gluon spectral function
by their respective tree-level expressions eqs.~\eqref{eq:qutprop} and \eqref{eq:gleqspectree},
\begin{multline}
\label{eq:1lquself}
[ i \Sigma_{\sigma}^{>,<} (q) ]_\text{one-loop} = - \frac{4 \gs^2}3 \int \frac{\text{d}^4 k}{(2\pi)^4} \left(2 g^{\sigma\alpha} g^{\beta\gamma} - g^{\sigma\gamma} g^{\alpha\beta} \right)
\\
\times [i S^{>,<}_{\gamma} (k)]_\text{tree} \, [i \Delta^{>,<}_{\alpha\beta} (q-k)]_\text{tree} \ .
\end{multline}
Inserting the expression in eqs.~\eqref{eq:gleqspectree} and \eqref{eq:1lquprop} into \eqref{eq:quark self energy},
and collecting all terms with a non-trivial Lorentz structure, one has to evaluate the object
\begin{align}
\label{eq:lorfact}
\frac1{q^4}  (2 q^\mu q^\sigma - q^2 g^{\mu\sigma}) 
Q_\mu (p,q) Q_\sigma(k,q) \ ,
\end{align}
where $p$ and $k$ are the momenta of the two quarks in the initial or final state, $q$ the exchanged momentum,
$(p-q)$ and $(k-q)$ the resulting gluon momenta.
The function
\begin{align}
Q^\mu (p,q) &\equiv ( 2 g^{\mu\rho} p^\nu - p^\mu g^{\nu\rho} ) \left(- g_{\nu\rho} +  (1- \alpha) \frac{(p-q)_\nu (p-q)_\rho}{(p-q)^2} \right)
\end{align}
depends on the gauge-fixing parameter $\alpha$.
This is to be expected because we have neglected various two-loop diagrams that capture interference terms,
and that would have to be included for a complete leading order computation of the collision term.
Therefore, the object given in~\eqref{eq:lorfact} can only be gauge parameter invariant at leading-logarithmic order,
i.e. after neglecting IR safe contributions and fixing the signs of the quark and gluon momenta such that $k_0 (k_0 - q_0) > 0$ and $(p_0 - q_0) > 0$.
This gives
\begin{align}
\frac1{q^4}  (2 q^\mu q^\sigma - q^2 g^{\mu\sigma}) Q^\mu (p,q) Q^\sigma(k,q) 
&\to - \frac{(pk)}{q^2} + \text{ IR safe} \ .
\end{align}
In this limit, inserting the Ansatz~\eqref{eq:devan}, and using the cutoff prescription~\eqref{eq:llcutoff}, one recovers the collision term
\begin{align}
\left[ \dot f_{fL} ( x,\vec p ) \right]_\text{qu}
&= - \frac{g_s^2 m_\infty^2 \ln(g_s^{-1})}{6\pi p_0}
(1+2 n_B(p_0)) n_F(p_0) (1-n_F(p_0)) \eta_{fL} (\bm p) \ ,
\end{align}
where a gain term that also depends on $\eta_{fL} (\vec p)$ vanishes
since the departure from equilibrium is odd under $\vec p \to - \vec p$.
This is the expression we use in eq.~\eqref{eq:collqqgg}, and it also is given in \cite{Ghiglieri:2018dib}.

\bibliography{references}

\providecommand{\href}[2]{#2}\begingroup\raggedright\begin{thebibliography}{10}

\bibitem{Bodeker:2020ghk}
D.~Bodeker and W.~Buchmuller, \emph{{Baryogenesis from the weak scale to the
  grand unification scale}},
  \href{http://dx.doi.org/10.1103/RevModPhys.93.035004}{\emph{Rev. Mod. Phys.}
  {\bf 93} (2021) 035004}, [\href{https://arxiv.org/abs/2009.07294}{{\tt
  2009.07294}}].

\bibitem{Moore:1998zk}
G.~D. Moore, \emph{{The Sphaleron rate: Bodeker's leading log}},
  \href{http://dx.doi.org/10.1016/S0550-3213(99)00746-4}{\emph{Nucl. Phys.}
  {\bf B568} (2000) 367--404},
  [\href{https://arxiv.org/abs/hep-ph/9810313}{{\tt hep-ph/9810313}}].

\bibitem{Moore:2010jd}
G.~D. Moore and M.~Tassler, \emph{{The Sphaleron Rate in SU(N) Gauge Theory}},
  \href{http://dx.doi.org/10.1007/JHEP02(2011)105}{\emph{JHEP} {\bf 02} (2011)
  105}, [\href{https://arxiv.org/abs/1011.1167}{{\tt 1011.1167}}].

\bibitem{Moore:2000mx}
G.~D. Moore, \emph{{Sphaleron rate in the symmetric electroweak phase}},
  \href{http://dx.doi.org/10.1103/PhysRevD.62.085011}{\emph{Phys. Rev.} {\bf
  D62} (2000) 085011}, [\href{https://arxiv.org/abs/hep-ph/0001216}{{\tt
  hep-ph/0001216}}].

\bibitem{Annala:2023jvr}
J.~Annala and K.~Rummukainen, \emph{{Electroweak sphaleron in a magnetic
  field}}, \href{http://dx.doi.org/10.1103/PhysRevD.107.073006}{\emph{Phys.
  Rev. D} {\bf 107} (2023) 073006},
  [\href{https://arxiv.org/abs/2301.08626}{{\tt 2301.08626}}].

\bibitem{Arnold:1999uy}
P.~B. Arnold and L.~G. Yaffe, \emph{{High temperature color conductivity at
  next-to-leading log order}},
  \href{http://dx.doi.org/10.1103/PhysRevD.62.125014}{\emph{Phys. Rev.} {\bf
  D62} (2000) 125014}, [\href{https://arxiv.org/abs/hep-ph/9912306}{{\tt
  hep-ph/9912306}}].

\bibitem{Silin:1960pya}
V.~P. Silin, \emph{{On the Electromagnetic Properties of a Relativistic
  Plasma}}, {\emph{Zh. Eksp. Teor. Fiz.} {\bf 38} (1960) 1577--1583}.

\bibitem{Blaizot:1992gn}
J.-P. Blaizot and E.~Iancu, \emph{{Kinetic theory and quantum electrodynamics
  at high temperature}},
  \href{http://dx.doi.org/10.1016/0550-3213(93)90490-G}{\emph{Nucl. Phys. B}
  {\bf 390} (1993) 589--620}.

\bibitem{Lifshitz1995Physical}
E.~Lifshitz and L.~Pitaevskii, \emph{Physical Kinetics}, vol.~10 of
  \emph{Course of Theoretical Physics}.
\newblock Elsevier Science, 1995.

\bibitem{Blaizot:1994am}
J.-P. Blaizot and E.~Iancu, \emph{{Energy momentum tensors for the quark -
  gluon plasma}},
  \href{http://dx.doi.org/10.1016/0550-3213(94)90517-7}{\emph{Nucl. Phys. B}
  {\bf 421} (1994) 565--592}, [\href{https://arxiv.org/abs/hep-ph/9401211}{{\tt
  hep-ph/9401211}}].

\bibitem{Blaizot:1993be}
J.~P. Blaizot and E.~Iancu, \emph{{Soft collective excitations in hot gauge
  theories}}, \href{http://dx.doi.org/10.1016/0550-3213(94)90486-3}{\emph{Nucl.
  Phys. B} {\bf 417} (1994) 608--673},
  [\href{https://arxiv.org/abs/hep-ph/9306294}{{\tt hep-ph/9306294}}].

\bibitem{Arnold:1996dy}
P.~B. Arnold, D.~Son and L.~G. Yaffe, \emph{{The Hot baryon violation rate is O
  (alpha-w**5 T**4)}},
  \href{http://dx.doi.org/10.1103/PhysRevD.55.6264}{\emph{Phys. Rev.} {\bf D55}
  (1997) 6264--6273}, [\href{https://arxiv.org/abs/hep-ph/9609481}{{\tt
  hep-ph/9609481}}].

\bibitem{Bodeker:1999ey}
D.~Bodeker, \emph{{From hard thermal loops to Langevin dynamics}},
  \href{http://dx.doi.org/10.1016/S0550-3213(99)00435-6}{\emph{Nucl. Phys. B}
  {\bf 559} (1999) 502--538}, [\href{https://arxiv.org/abs/hep-ph/9905239}{{\tt
  hep-ph/9905239}}].

\bibitem{Bodeker:1998hm}
D.~Bodeker, \emph{{On the effective dynamics of soft nonAbelian gauge fields at
  finite temperature}},
  \href{http://dx.doi.org/10.1016/S0370-2693(98)00279-2}{\emph{Phys. Lett.}
  {\bf B426} (1998) 351--360},
  [\href{https://arxiv.org/abs/hep-ph/9801430}{{\tt hep-ph/9801430}}].

\bibitem{Arnold:2000dr}
P.~B. Arnold, G.~D. Moore and L.~G. Yaffe, \emph{{Transport coefficients in
  high temperature gauge theories. 1. Leading log results}},
  \href{http://dx.doi.org/10.1088/1126-6708/2000/11/001}{\emph{JHEP} {\bf 11}
  (2000) 001}, [\href{https://arxiv.org/abs/hep-ph/0010177}{{\tt
  hep-ph/0010177}}].

\bibitem{Ghiglieri:2018dib}
J.~Ghiglieri, G.~D. Moore and D.~Teaney, \emph{{QCD Shear Viscosity at (almost)
  NLO}}, \href{http://dx.doi.org/10.1007/JHEP03(2018)179}{\emph{JHEP} {\bf 03}
  (2018) 179}, [\href{https://arxiv.org/abs/1802.09535}{{\tt 1802.09535}}].

\bibitem{Arnold:2002ja}
P.~B. Arnold, G.~D. Moore and L.~G. Yaffe, \emph{{Photon and gluon emission in
  relativistic plasmas}},
  \href{http://dx.doi.org/10.1088/1126-6708/2002/06/030}{\emph{JHEP} {\bf 06}
  (2002) 030}, [\href{https://arxiv.org/abs/hep-ph/0204343}{{\tt
  hep-ph/0204343}}].

\bibitem{Arnold:2003zc}
P.~B. Arnold, G.~D. Moore and L.~G. Yaffe, \emph{{Transport coefficients in
  high temperature gauge theories. 2. Beyond leading log}},
  \href{http://dx.doi.org/10.1088/1126-6708/2003/05/051}{\emph{JHEP} {\bf 05}
  (2003) 051}, [\href{https://arxiv.org/abs/hep-ph/0302165}{{\tt
  hep-ph/0302165}}].

\bibitem{Heiselberg:1994vy}
H.~Heiselberg, \emph{{Viscosities of quark - gluon plasmas}},
  \href{http://dx.doi.org/10.1103/PhysRevD.49.4739}{\emph{Phys. Rev. D} {\bf
  49} (1994) 4739--4750}, [\href{https://arxiv.org/abs/hep-ph/9401309}{{\tt
  hep-ph/9401309}}].

\bibitem{Caron-Huot:2008zna}
S.~Caron-Huot, \emph{{O(g) plasma effects in jet quenching}},
  \href{http://dx.doi.org/10.1103/PhysRevD.79.065039}{\emph{Phys. Rev. D} {\bf
  79} (2009) 065039}, [\href{https://arxiv.org/abs/0811.1603}{{\tt
  0811.1603}}].

\bibitem{Bodeker:2000da}
D.~Bodeker, \emph{{A Local Langevin equation for slow long distance modes of
  hot nonAbelian gauge fields}},
  \href{http://dx.doi.org/10.1016/S0370-2693(01)00911-X}{\emph{Phys. Lett. B}
  {\bf 516} (2001) 175--182}, [\href{https://arxiv.org/abs/hep-ph/0012304}{{\tt
  hep-ph/0012304}}].

\bibitem{Bodeker:2002gy}
D.~Bodeker, \emph{{Perturbative and nonperturbative aspects of the nonAbelian
  Boltzmann-Langevin equation}},
  \href{http://dx.doi.org/10.1016/S0550-3213(02)00841-6}{\emph{Nucl. Phys. B}
  {\bf 647} (2002) 512--538}, [\href{https://arxiv.org/abs/hep-ph/0205202}{{\tt
  hep-ph/0205202}}].

\bibitem{Gunion:1989we}
J.~F. Gunion, H.~E. Haber, G.~L. Kane and S.~Dawson, \emph{The Higgs Hunter's
  Guide}.
\newblock CRC Press, 1990,
  \href{http://dx.doi.org/10.1201/9780429496448}{10.1201/9780429496448}.

\bibitem{ParticleDataGroup:2024cfk}
{\scshape Particle Data Group} collaboration, S.~Navas et~al., \emph{{Review of
  particle physics}},
  \href{http://dx.doi.org/10.1103/PhysRevD.110.030001}{\emph{Phys. Rev. D} {\bf
  110} (2024) 030001}.

\bibitem{Arnold:1998cy}
P.~B. Arnold, D.~T. Son and L.~G. Yaffe, \emph{{Effective dynamics of hot, soft
  nonAbelian gauge fields. Color conductivity and log(1/alpha) effects}},
  \href{http://dx.doi.org/10.1103/PhysRevD.59.105020}{\emph{Phys. Rev. D} {\bf
  59} (1999) 105020}, [\href{https://arxiv.org/abs/hep-ph/9810216}{{\tt
  hep-ph/9810216}}].

\bibitem{Prokopec:2003pj}
T.~Prokopec, M.~G. Schmidt and S.~Weinstock, \emph{{Transport equations for
  chiral fermions to order h bar and electroweak baryogenesis. Part 1}},
  \href{http://dx.doi.org/10.1016/j.aop.2004.06.002}{\emph{Annals Phys.} {\bf
  314} (2004) 208--265}, [\href{https://arxiv.org/abs/hep-ph/0312110}{{\tt
  hep-ph/0312110}}].

\bibitem{Prokopec:2004ic}
T.~Prokopec, M.~G. Schmidt and S.~Weinstock, \emph{{Transport equations for
  chiral fermions to order h-bar and electroweak baryogenesis. Part II}},
  \href{http://dx.doi.org/10.1016/j.aop.2004.06.001}{\emph{Annals Phys.} {\bf
  314} (2004) 267--320}, [\href{https://arxiv.org/abs/hep-ph/0406140}{{\tt
  hep-ph/0406140}}].

\bibitem{Berges:2004yj}
J.~Berges, \emph{{Introduction to nonequilibrium quantum field theory}},
  \href{http://dx.doi.org/10.1063/1.1843591}{\emph{AIP Conf. Proc.} {\bf 739}
  (2004) 3--62}, [\href{https://arxiv.org/abs/hep-ph/0409233}{{\tt
  hep-ph/0409233}}].

\bibitem{Drewes:2016gmt}
M.~Drewes, B.~Garbrecht, D.~Gueter and J.~Klaric, \emph{{Leptogenesis from
  Oscillations of Heavy Neutrinos with Large Mixing Angles}},
  \href{http://dx.doi.org/10.1007/JHEP12(2016)150}{\emph{JHEP} {\bf 12} (2016)
  150}, [\href{https://arxiv.org/abs/1606.06690}{{\tt 1606.06690}}].

\bibitem{Garbrecht:2018mrp}
B.~Garbrecht, \emph{{Why is there more matter than antimatter? Calculational
  methods for leptogenesis and electroweak baryogenesis}},
  \href{http://dx.doi.org/10.1016/j.ppnp.2019.103727}{\emph{Prog. Part. Nucl.
  Phys.} {\bf 110} (2020) 103727},
  [\href{https://arxiv.org/abs/1812.02651}{{\tt 1812.02651}}].

\bibitem{Elze:1989ba}
H.-T. Elze, \emph{{SELFCONSISTENT ONE LOOP TRANSPORT EQUATIONS FOR QCD}},  in
  \emph{{NATO ASI: International Advanced Courses on The Nuclear Equation of
  State}}, 6, 1989.

\bibitem{Cornwall:1974vz}
J.~M. Cornwall, R.~Jackiw and E.~Tomboulis, \emph{{Effective Action for
  Composite Operators}},
  \href{http://dx.doi.org/10.1103/PhysRevD.10.2428}{\emph{Phys. Rev. D} {\bf
  10} (1974) 2428--2445}.

\bibitem{Braaten:1989mz}
E.~Braaten and R.~D. Pisarski, \emph{{Soft Amplitudes in Hot Gauge Theories: A
  General Analysis}},
  \href{http://dx.doi.org/10.1016/0550-3213(90)90508-B}{\emph{Nucl. Phys. B}
  {\bf 337} (1990) 569--634}.

\bibitem{Gorda:2023zwy}
T.~Gorda, R.~Paatelainen, S.~S{\"a}ppi and K.~Sepp{\"a}nen, \emph{{Soft gluon
  self-energy at finite temperature and density: hard NLO corrections in
  general covariant gauge}},
  \href{http://dx.doi.org/10.1007/JHEP08(2023)021}{\emph{JHEP} {\bf 08} (2023)
  021}, [\href{https://arxiv.org/abs/2304.09187}{{\tt 2304.09187}}].

\end{thebibliography}\endgroup
\bibliographystyle{jhep}

\end{document}